# Extremely elastic soliton crystals generated in a passively mode-locked tunable high-repetition-rate fiber laser

ALEXEY ANDRIANOV,* ARKADY KIM

*Institute of Applied Physics of the Russian Academy of Sciences, Nizhny Novgorod, Russia*
*\*Corresponding author: alex.v.andrianov@gmail.com*

We present the first direct observation of the bound state of multiple dissipative optical solitons in which bond length and bond strength can be individually controlled in a broad range in a regular manner. We have observed experimentally a new type of stable and extremely elastic soliton crystals that can be stretched and compressed many times conserving their structure by adjusting the bond properties in real time in a specially designed passively mode-locked fiber laser incorporating highly asymmetric tunable Mach-Zehnder interferometer. The temporal structure and dynamics of the generated soliton crystals have been studied using an asynchronous optical sampling system with picosecond resolution. We demonstrated that stable and robust soliton crystal can be formed by two types of primitive structures: single dissipative solitons, and(or) pairs of dissipative soliton and pulse with lower amplitude. Continuous stretching and compression of a soliton crystal with extraordinary high ratio of more than 30 has been demonstrated with a smallest recorded separation between pulses as low as 5 ps corresponding to an effective repetition frequency of 200 GHz. Collective pulse dynamics, including soliton crystal self-assembling, cracking and transformation of crystals comprising pulse pairs to the crystals of similar pulses has been observed experimentally.

## 1. INTRODUCTION

Being a universal concept dissipative solitons are encountered in various dynamical systems making a bridge between different fields of science [1]. Even a deeper interdisciplinary connection can be established since it was shown that large ensembles of dissipative solitons demonstrate some thermodynamical properties and phase transitions between "soliton gas", "soliton liquid" and "soliton crystal" [2]. The ability of dissipative solitons to form bound states of multiple pulses, which are also called "soliton molecules" (SMs) or "soliton crystals" (SCs) [3] highlighting the analogy with the physics of matter, attracts much attention due to its fundamental importance for collective dynamics of dissipative nonlinear systems and possible practical applications.

One of the most explored areas where dissipative solitons and soliton molecules naturally appear is the field of mode-locked lasers and, especially, ultrashort fiber lasers [4]. Generation of stable optical SMs was demonstrated in a variety of laser cavity configurations and was confirmed by numerical modeling and supported by analytical studies [3]. Optical SMs can be used as information carriers in ultra-high-speed transmission systems, for optical storage and for many other practical applications.

The basic mechanism underlying the formation of SMs relies on the interaction of individual solitons, which results in one or more stable configurations of the molecule with different bond lengths (i.e. a separation of solitons) [3,5,6]. To date, most studies consider the soliton-to-soliton bond formation cause by soliton interaction resulting from overlapping of the solitons and their neighbor's wings that are formed due to pulse shaping, given the balance of nonlinearity and dispersion. Thus, very limited control of the bond properties is possible, since variations of nonlinearity or dispersion lead to dramatic changes in the system dynamics.

The research on optical SMs generation and their properties has a significant overlap with the studies aimed at building mode-locked lasers with extremely high repetition rate reaching the multi-hundred gigahertz and even terahertz range. Pulse packets and trains with a very small spacing between pulses starting from a few picoseconds are very promising for applications in high-speed data transmission and processing systems, for frequency comb generation, and ultrafast sampling of optical as well as electrical signals. Several approaches to the generation of such pulses were proposed and tested experimentally, from which we highlight lasers with intracavity comb-like frequency filters and lasers generating bunches of soliton pulses that were interpreted as soliton molecules [7].

Mode-locked lasers with intracavity spectral filters such as a Fabry-Perrot (FP) filter [8, 9], high finesse microcaviy [10] or

Mach-Zehnder interferometer (MZI) [11, 12] were reported to generate coherent pulse sequences where the separation between the pulses is governed by the inverse frequency difference between the filter lines. Generation of continuous pulse trains was demonstrated in the laser with FP filter and microcavity, however, recent study [13] showed that unstable and irregular pulse patterns are likely to form in the laser with intracavity MZI. Particular mode-locking mechanisms responsible for high repetition rate pulse trains formation are still discussed. The effect of intracavity comb-like spectral filter can be interpreted in the time domain as a non-uniform response function having a more or less periodic structure with the period inversely proportional to the frequency separation of spectral comb teeth. This is clearly seen for unbalanced MZI, whose output is a coherent combination of the input signal and its delayed replica. An intracavity element with such a response may have a significant impact on the generation of soliton crystals by providing a new mechanism for the soliton-to-solton bond formation with externally controllable properties. Despite the numerous studies on mode-locked lasers with intracavity filters there was no clear evidence of the possibility of generation of soliton crystals governed by the filter response. We note, that the systems with delayed response that may lead to formation of bound state of disspiative solitons were studied in the context of vertical cavity emitting laser placed in an external double cavity [14, 15].

A systematic study of the dynamics of high-repetition-rate irregular pulse sequences in experiment is challenging because continuous recording of ultrafast shapes with picosecond resolution within a time window corresponding to the round-trip of the cavity, which is usually of the order of several nanoseconds, is generally required. Direct observations of the pulse sequences using an ultrafast photodetector and a sampling oscilloscope were presented in [5,16]; however, picosecond transients with THz bandwidth are still inaccessible for electronic oscilloscopes. Well-established methods for reconstructing ultrashort pulse shapes utilizing nonlinear optical interactions such as autocorrelation, FROG and similar methods possess femtosecond resolution but a limited temporal window size rarely exceeding 100 ps and slow update rate. Another complication is low energy of high-repetition-rate pulses, usually well below 1 nJ, making single-shot nonlinear measurements difficult. As a multiple time scale dynamics involving slow dynamics in the mode-locked lasers may come into play [16, 17], averaged measurements become generally unacceptable. To date, the majority of works studying the ultrafast temporal structure of the pulse sequences generated in mode-locked lasers utilize autocorrelation measurements which provide very limited information on a particular sequence shape and cannot track temporal dynamics. Recently developed methods based on the time lens proved their ability to capture complex optical waveforms in real-time in single-shot regime and were used to study transient dissipative soliton dynamics in a mode-locked laser [18, 19] Simpler technique utilizing spectral interferometry and time-stretch dispersive Fourier-transform was used for probing the internal dynamics of an optical soliton molecule [20]. However, limited temporal window size provided by time lensing and spectral interferometry, which is far smaller than the cavity roundtrip time, makes it difficult to use these methods to measure long pulse sequences.

In this context, the technique known as asynchronous optical sampling appears very promising since the compromise between temporal resolution, temporal window size, sensitivity and acquisition rate can be achieved in a simple setup. Recently optical sampling was successfully applied to measure high-speed pseudo-random optical data streams with subpicosecond resolution [21-23].

In this article we present the first direct observation of the bound state of multiple dissipative solitons in which the bond properties, namely bond length and bond strength, can be individually controlled in a very broad range in a regular manner. This enables generation of a new type of stable and extremely elastic soliton crystals that can be stretched and compressed many times conserving their structure by adjusting the bond properties in real time. Soliton crystals are produced in a specially designed passively mode-locked fiber laser incorporating highly asymmetric tunable MZI. The temporal structure and dynamics of the generated soliton crystals was studied using an asynchronous optical sampling system with picosecond resolution. Continuous stretching and compression of the soliton crystal conserving the number of pulses with extraordinary high ratio of more than 30 was demonstrated with a smallest recorded pulse separation as low as 5 ps corresponding to an effective repetition frequency of 200 GHz. Experiments reveal that elastic soliton crystals exhibit some properties unusual for ordinary soliton crystals such as coexistence of two stable configurations of pulses, which can be transformed one into the other.

## 2. EXPERIMENTAL SETUP AND OBSERVATION OF ELASTIC SOLITON CRYSTALS

The schematic of the fiber laser generating elastic stretchable soliton crystals (SC laser) is shown in Fig. 1. The laser consists of three main parts: (I) a nonlinear amplifying loop mirror (NALM) acting as an artificial saturable absorber, (II) a propagation loop incorporating an Er-doped fiber amplifier for cavity loss compensation, and (III) a highly asymmetric unbalanced MZI formed by two couplers, one of which also directs part of radiation to the laser output, and an adjustable optical delay line [24]. The fibers are non-polarization-maintaining (non-PM), so the polarization controllers are installed.

A 3-meter piece of highly nonlinear fiber with small anomalous dispersion (-13ps$^2$/km) is incorporated into the NALM to increase nonlinear interaction and lower the mode-locking threshold, thus allowing laser operation with many pulses in the cavity keeping the average intracavity power at a moderate level of 300 mW. Active fibers in the NALM and the propagation loop have the same length of 1.4 m and normal dispersion (23 ps$^2$/km). The other fibers forming the cavity are standard single-mode fibers with anomalous dispersion. The laser is pumped by two 600 mW single-mode laser diodes at 980 nm. The total length of the laser cavity is 7.8 m resulting in the fundamental repetition frequency of 25.04 MHz. A NALM fiber splitter has coupling ratio 80/20 unlike the commonly used 50/50, which results in lower unsaturated losses of artificial saturable absorber, but modulation depth sufficient to achieve mode-locking.

The most important part of the laser is the asymmetric unbalanced MZI incorporated into the propagation loop that injects a delayed replica of the signal back into the cavity with adjustable delay and amplitude. A home-built delay line is formed by an optical circulator, collimator and beam reverser mounted on a translation stage. In contrast to the previous studies on high-repetition-rate mode-locked lasers utilizing MZI with 50/50 couplers, we used 80/20 couplers and a variable attenuator in the free-space path so as only about 0-5% of the delayed signal was returned to the cavity.

The net cavity dispersion is low anomalous (-0.023 ps$^2$) which, combined with high nonlinearity and high pump power, provides conditions sufficient for the generation of multiple dissipative solitons per roundtrip.

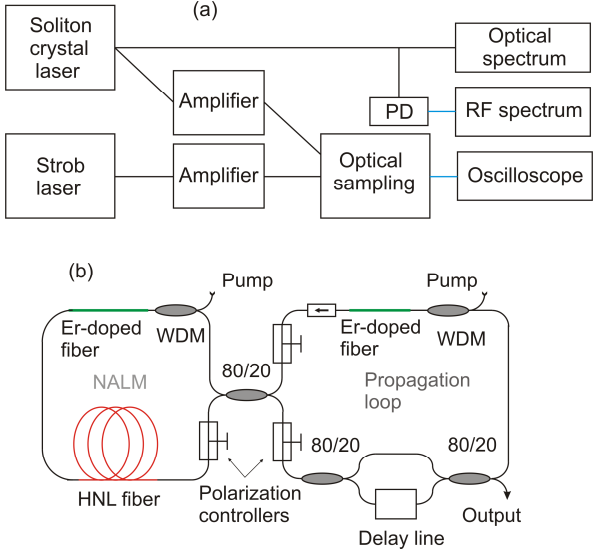

Fig. 1. (a) Overview of the experiment on the generation of "elastic soliton crystals", (b) schematic of the fiber laser.

The energy of the dissipative soliton is known to be quantized so that at high enough pump power a large number of pulses can simultaneously circulate and interact inside the cavity. The properties of the pulse sequences generated in mode-locked lasers were studied in many works. It was shown that in the laser without intracavity MZI the solitons can condensate into regular stable structures in which the separation between pulses is essentially determined by the interaction of the pulse wings. However, when an unbalanced MZI is incorporated into the cavity, it provides a delayed replica of the pulse, and the next pulse can be anchored to the position precisely determined by the MZI delay via nonlinear interaction with that replica rather than with the tail of the previous pulse, as is shown schematically in Fig. 2a. Moreover, the next pulse can track slow variations of the delay. Several dissipative solitons assembled into a chain in which each pulse (except for the first one) interacts with the delayed replica of the previous one, form a new stable multisoliton state which we named "elastic soliton crystal".

To measure the temporal structure of the generated pulse sequence we implemented the asynchronous optical sampling technique in the setup shown in Fig. 1. To generate sampling pulses another fiber laser mode-locked via nonlinear polarization rotation in the single pulse regime at the fundamental frequency was built. By tailoring the cavity length the repetition rate of the sampling laser was adjusted to the value about 600 Hz higher than the fundamental frequency of the soliton crystal laser. Sampling pulses were amplified in an Er-doped fiber amplifier and drove the optical sampling system based on an all-fiber nonlinear loop mirror. Based on stroboscopic optical sampling it continuously acquired an intensity profile at the laser output with ~5 ps resolution within the round-trip time window (39.9 ns) with the rate of 600 frames per second [25].

After switching on the pump of the laser mode-locked operation was self-started with several pulse complexes and individual pulses randomly distributed over the round-trip period. It was found that to generate the most ordered sequences it was sufficient to adjust the MZ delay $T$ from negative to positive (or from positive to negative) crossing the zero delay $T=0$ with the speed of the order of 1ps/1s. At small delays ($T<5$ ps) the laser switched to the chaotic generation of individual pulses and pulse clusters resembling the scenario commonly referred to as "soliton rain" [26]. Pulses and pulse clusters repetitively emerged from noise and drifted, sometimes condensing into unstable bunches, which could randomly dissociate into sub-bunches and even disappear. The optical spectrum at the laser output in this regime was unstable and represented a mixture of spectra of individual pulses. However, as the delay line moved towards higher $T$ (positive or negative), a well-ordered sequence of pulses emerged with the pulse separation equal to the delay line setting.

Usually the pulses are grouped into one or several stable bunches. The separation between pulses in the bunch is equal to the delay $T$ (to within the calibration error of about 0.2 ps) and the bunch length is $nT$, where $n$ is the number of pulses in the bunch. These stable pulse clusters are considered to be a new type of a multisoliton bound state, which can be called "elastic soliton crystal". The fundamental difference from conventional "soliton crystals" is that the interaction between neighboring pulses is governed by the pulse replicas superimposed by the MZ interferometer rather than by the pulse tails. This allows easy control of the bond parameters such as bond strength and bond length. The dissipative solitonic nature of the generated pulses manifests itself in the uniformity of the pulse parameters and quantization of the total energy circulating in the cavity, which is directly proportional to the number of generated pulses.

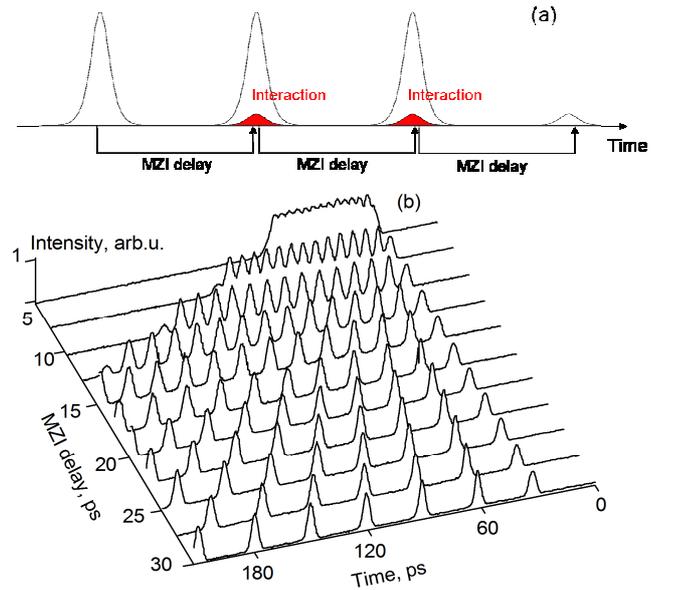

Fig. 2. (a) Mechanism of elastic soliton crystal formation, (b) continuous compression of elastic soliton crystal comprising 14 solitons recorded in experiment by optical sampling setup while adjusting MZI delay.

The experimentally observed stretching and compression of the soliton crystal is presented in Fig. 2b. On formation of a stable SM the MZI delay line is moved back and forth several times. The temporal shapes recorded by optical sampling clearly show that the separation between pulses follows the delay line setting. The number of pulses in the molecule is conserved. The smallest observed separation between pulses at which the molecule is stable is about 5 ps, corresponding to the effective repetition rate of 200 GHz. We believe that one main reason for the instability appearing at smaller separations is the competition of the interactions of the next pulse in the sequence with the delayed replica and with the previous pulse wing, both of which are phase sensitive and in a coherent combination are not robust to slight phase drifts inevitable in the system. The largest recorded separation between the solitons was 0.29 ns, corresponding to 3.4 GHz repetition rate, and was limited by the delay line travel range. Thus, the elastic soliton crystal can be stretched and compressed by a factor of about 60. Interestingly, with large separation the soliton crystal completely fills the cavity and can be locked to itself resulting in the generation of a continuous pulse train. The transition of the single crystal to the continuous pulse train is shown in Fig. 3. In this regime the resulting repetition rate is exactly equal to the high harmonic of the fundamental cavity frequency. It is interesting to note that unlike the recently reported mode-locked laser with MZI [12,13], the generated high-repetition-rate sequence of solitons in our experiment was quite stable. The optical spectrum in this regime shown in Fig. 3c is formed by a very broad envelope corresponding to the spectrum of the individual pulse and fast oscillations with a period inversely proportional to the pulse separation.

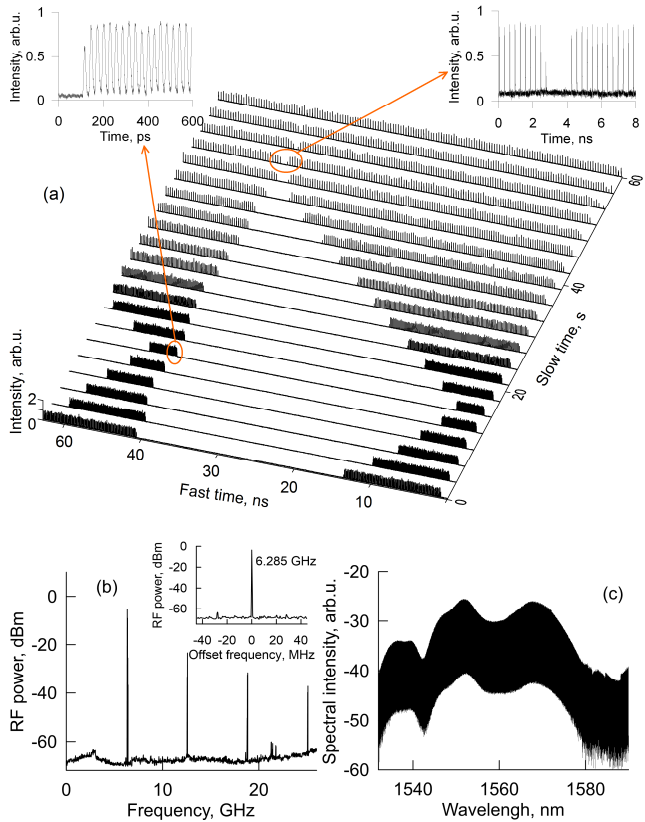

Fig. 3. (a) Compression and stretching of 136-soliton crystal which finally fills the entire cavity resulting in the generation of continuous pulse train with repetition rate 3.4 GHz, (b) RF spectrum of 251-soliton crystal, (c) optical spectrum at the laser output. The insets show magnified pulse patterns at highest compression point and immediately before filling the entire cavity.

The radio-frequency spectrum of 251 soliton crystal measured with a 40 GHz photodiode and a spectrum analyzer clearly shows narrow line generation at as high as 251th harmonic (6.3 GHz) with side-bands suppression higher than 50 dB (see Fig. 3b). We also notice the possibility of soliton annihilation when we tried to gradually stretch the soliton crystal more than the cavity round-trip time, i.e. "overfill" the cavity. In this case the soliton at the leading edge of the crystal inevitably meets the soliton at its trailing edge, and the process usually ends up with annihilation of the solitons during interaction. The excess intracavity energy maintained by the laser pump then emerges as a CW component, which can be seen in the optical spectrum as a narrow line. In this way we can precisely control the harmonic number, and also obtain soliton crystals with a different number of pulses.

## 3. ELASTIC SOLITON CRYSTAL OF DIFFERENT COMPOUNDS

We observed that stable pulse complexes assembled from several different types of pulses distinguished by their amplitudes can be generated in the laser simultaneously. Such complexes behave as a single elastic soliton crystal, i.e. can be stretched and compressed by adjusting the MZI while preserving their structure. An example of such a crystal and its compression is shown in Fig. 4a. Careful examination of many recordings taken with our optical sampling system reveals that there are two basic structures that can be assembled into a crystal: a single high-amplitude pulse (H-pulse) and a pair of high- and low-amplitude pulses (H-L pulse pairs). However, we never observed two low-amplitude pulses following each other. Based on this observation and numerical simulations we conclude that the H-L pulse pair is actually a combination of a dissipative soliton (H-pulse) followed by its delayed replica (L-pulse) as is shown schematically in Fig. 4b. The amplitude of the L-pulse is insufficient to trigger the formation of a soliton, but sufficient to make one more round-trip in the cavity and produce a delayed replica in the MZI serving as an anchoring point for the next structure that can be a single soliton or a pair as well. For correct interpretation of measured pulse shape we remind the reader that in our highly asymmetric MZI the larger part of the delayed replica (80%) is directed to the laser output.

Formation of stable irregular soliton patterns, which are sometimes referred to as macromolecules, with different separations between pulses examined by fast photodetector and oscilloscope was reported [4-6]. Coexistence of two different types of dissipative solitons and switching between them was also demonstrated [27]. However, generation of bound states of pulses with notably different amplitudes in mode-locked lasers has not been well studied.

At some carefully adjusted position of polarization controllers and the attenuation in a free-space path of MZI an elastic soliton crystal consisting only of pulse pairs can be generated routinely. Such a crystal can be stretched to fill all the cavity and can be locked to itself, thus providing at the output a stable continuous train of interleaving pulses. The stretching ratio of observed interleaved soliton crystal is about 30. It is noteworthy that the total number of pulses per roundtrip that can be generated in this unusual regime of high-order harmonic mode-locking is two times larger (to the margin of few pulses) than the number of equal solitons that can be generated at the same pump power. This observation further supports our scenario of formation of the pulse pairs, in which the main amount of energy inside the cavity is carried by solitons interleaved with small pulses. The RF spectrum of the interleaved pulse train shows the main line at even harmonics of the resonator, which was as high as 2×250, corresponding to the frequency of 12.52 GHz. As expected, sub-harmonic line at 6.26 GHz with 10 dB smaller power was also observed [28].

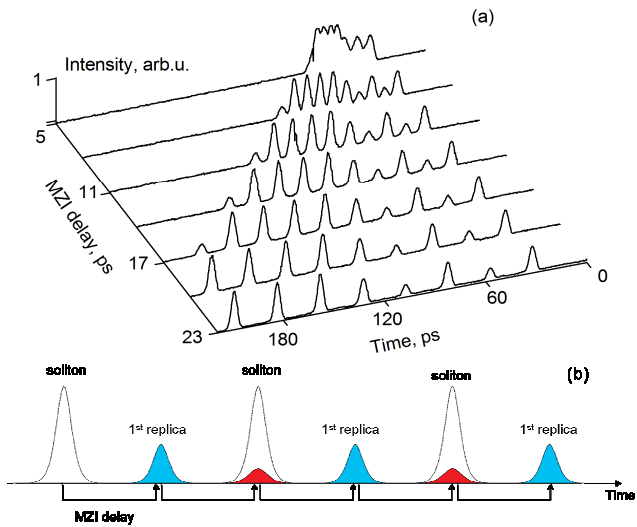

Fig. 4 (a) Compression of patterned soliton crystal comprising 2 pairs and 5 consecutive solitons; (b) mechanism of formation of interleaved H-L pulse pattern.

We believe that the key property of our laser, which allows formation and coexistence of two phases in the soliton crystal, is a large and controllable asymmetry of the MZI splitting ratio. Careful choice of the amount of the delayed pulse replica injected back into the cavity and relatively high unsaturated transmission of NALM in contrast to the previously reported schemes enables the generation of such patterned elastic soliton crystals.

## 4. SELF-ASSEMBLING OF ELASTIC SOLITON CRYSTALS, CRYSTAL CRACKING AND PHASE TRANSFORMATIONS

Slow pulse dynamics in mode-locked fiber lasers was found to have a strong impact on the generation and interaction of multiple pulse bunches [16]. Characteristic time scales on which slow changes occur lie in the millisecond range and even larger, which is much slower than usual cavity roundtrip time. Mechanisms responsible for these variations are usually attributed to the population inversion dynamics in an active medium and interaction with low-level coherent background; however, many aspects of the underlying processes have not been explored yet. In a laser generating multiple pulse bunches per round-trip, slow dynamics usually leads to relative movement of initially separated structures, which interact via fast nonlinear effects when they are close enough. It was shown that slow dynamics may assist self-assembling of dissipative soliton crystals [4].

In our laser slow dynamics has a significant impact on the formation of bound states and their further propagation in the cavity.

Our optical sampling system with 600 Hz update rate permits continuous monitoring of the pulse sequence over the round-trip period and resolving individual pulses virtually in real-time. The observed slow dynamics was most pronounced after the perturbation of the cavity by rapidly changing the MZI delay. An example of the evolution of the pulse sequence after moving the delay line is shown in Fig. 5. Close inspection of the recorded traces shows that at the beginning of the process the pulse sequence comprises many fragments consisting of H-pulse solitons or interleaving H-L pulse pairs. Temporal spacing between pulses in the fragments follows the delay line setting. When the delay line is stopped, the fragments continue moving slowly (the characteristic speed is a few ns/s) and can finally assemble into one or several stable large elastic soliton crystals. If the delay line is shaken again, the process repeats. We noticed that frequently the crystal parts are assembled back into the original structure.

One of the most intriguing features of the elastic soliton crystal dynamics is the possibility of transformation of one basic structure into the other. By careful examination of recorded traces and counting the total number of pulses we found that the H-L pulse pair can transform into a single H-pulse soliton. Transformation can be triggered by variations of the MZI delay or interaction of colliding soliton crystals, but rarely occurs spontaneously. The tendency to transformation strongly depends on settings of polarization controllers and MZI losses in the delayed arm; however, the uniform structures of the same H-pulse solitons are usually more stable. The total number of the pairs plus single solitons is conserved during the transformation. This observation supports our explanation of the formation of H-L pulse pairs.

In view of the analogy usually drawn between "soliton molecules" or "soliton crystals" and real molecules or crystals, pattern formation and coexistence of two different basic structures can be interpreted in two ways. The first way, inspired by the previous research [3, 29], implies the process of formation of a patterned soliton crystal as "copolymerization" of two types of "monomers", namely, a H-pulse soliton and a pair of H-L pulses. However, this picture does not reflect transformations of one "soliton monomer" into the other, which is not a usual case for real world molecules.

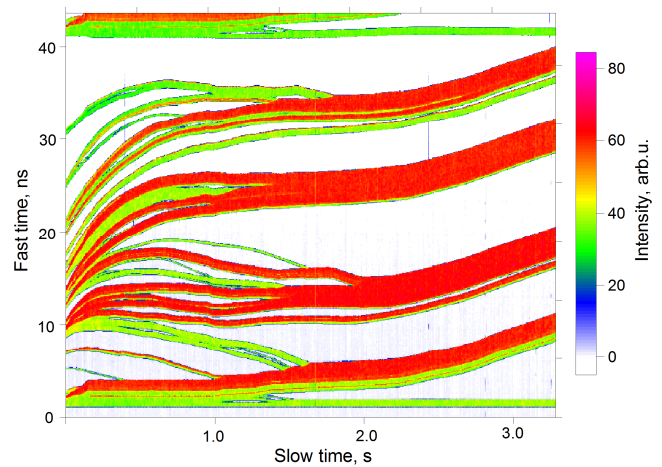

Fig. 5. Evolution of a pulse sequence containing many fragments of elastic soliton crystal after perturbing the cavity. Individual pulses are not resolved in the figure, however, H-L pulse pairs and H-pulses can be distinguished by their intensity: pulse pairs have about two times lower intensity (green color) than H-pulse solitons (red color).

The second way that we suggest interprets two different basic structures as two stable phases in 1D soliton crystalline structure, which can coexist and transform one into the other. The so-called solid-solid phase transformation [30] has attracted much attention recently in the material science. The real-world examples are some memory-shape materials or super-elastic materials, e.g. nickel-titanium alloy (nitinol) and the recently demonstrated organic crystals [31]. Their crystalline structure allows two stable coexisting phases (the so-called parent and daughter phases), the transformation between them occurs as the result of external action like changing the temperature or application of mechanical stress. Most importantly, the dimensions of a unit cell in such structures change significantly during the transformation. Similar behavior is observed for soliton crystals in our fiber laser, where the unit cell shrinks twice during the phase transformation. However, in our experiment the transformation of a single soliton to a pulse pair was observed very rarely, while transformation of pairs to single pulses was quite common.

The behavior observed in our fiber laser highlights the unique properties of elastic soliton crystals that manifest themselves in coexistence of two soliton crystalline phases.

## 5. NUMERICAL MODELING

For verification of our experimental findings we performed numerical modeling of the mode-locked laser with intracavity MZI. Comprehensive numerical simulation in a distributed model fully taking into account nonlinear propagation and polarization dynamics of pulses in the laser scheme and search for the parameter range within which mode-locking is supported is very time-consuming. We note, that numerical study of the laser incorporating an MZI was presented recently [32], however, no clear evidence of soliton crystals formation was reported. We have developed a highly simplified model with lumped parameters and fixed linear polarization (see Supplement 1 for details), which, however, reflects the main features of the observed laser dynamics. Indeed, with a proper choice of model parameters, formation of a bunch of equispaced dissipative solitons from initial noise was observed.

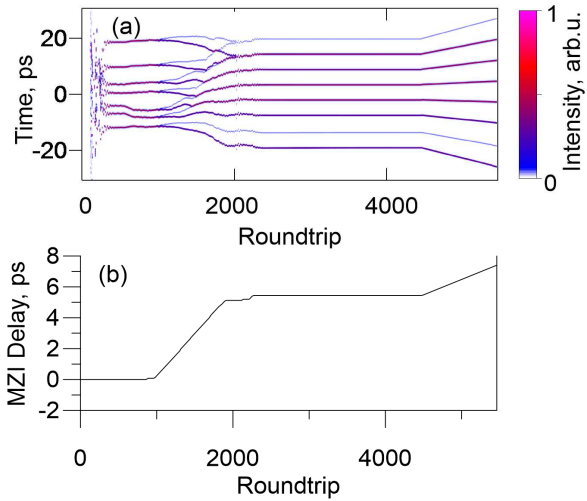

Fig. 6. Simulation of elastic soliton crystal formation: evolution of the pulse pattern in the cavity (a) and MZI delay (b).

Figure 6 shows the formation of a soliton crystal in the situation resembling our experiment. The simulation started from noise in the cavity and zero delay of the MZI. After a short time a few solitons randomly distributed in the cavity are formed. Then we manually adjusted the delay while monitoring the pulse pattern in the cavity in real time. At some point most of the pulses in the cavity become locked to the delayed replicas of the previous pulses and an elastic soliton crystal is formed. The separation of the pulses follows the setting of the delay that was maintained constant for some time and then increased slowly just to demonstrate the stability and tunability of the elastic soliton crystal. We found that the proposed starting procedure is important for the formation of a long soliton crystal. Starting the laser at some fixed delay led to the formation of individual pulses that were very rarely locked to each other, but could be finally assembled into a soliton crystal by adjusting the delay back and forth several times.

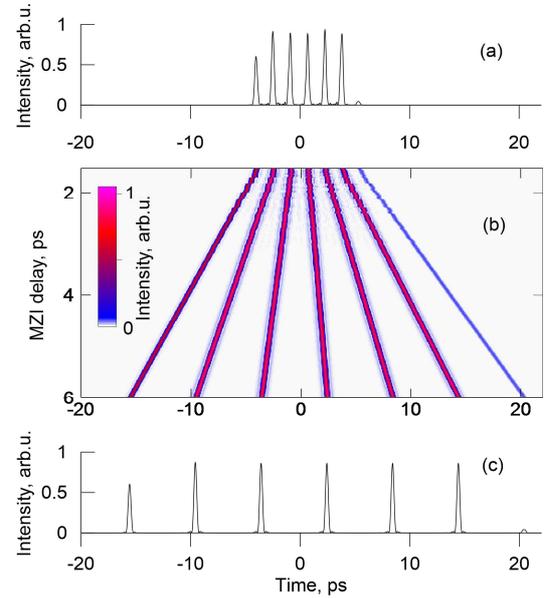

Fig. 7. Compression of soliton crystal comprising 6 solitons: (a) intensity profile after compression, evolution of intensity profile during compression (b), intensity profile before compression (c).

The simulation of a soliton crystal is shown in Fig. 7. It is seen that the pulse separation smoothly tunes down to about 3 ps following the delay setting. At smaller delays the pulse separation changes with glitches caused by the interaction of the pulse wings. The process of soliton molecule compression usually ends up at delays ~2 ps with soliton crystal fracture into several parts or annihilation of a few pulses.

By carefully adjusting the model parameters we observed the formation of an SC that consisted of the pulse pairs in which H-pulses were interleaved with L-pulses (see Fig. 8). In numerical experiments we slightly changed the amount of the delayed pulse replicas injected back into the cavity by MZI, i.e. changed the ratio $\delta=P_d/P$, where $P$ and $P_d$ are the powers of the main pulse and its delayed replica, respectively. We found that SCs formed by identical pulses more frequently appear at lower $\delta\sim0.02$, while SCs formed by interleaved H-L pulses more frequently appear at slightly higher $\delta\sim0.05$.

Careful examination of the pulse pair revealed that the H-pulse in the interleaved sequence is indeed the dissipative soliton maintained by the balance of linear and nonlinear loss and gain, while L-pulses are their delayed replicas. We verified that the intensity of the L-pulses is insufficient to pass through a saturation absorber with small enough losses that can be fully compensated by the gain. However, the L-pulses make one more round-trip in the cavity and their delayed replicas produced by MZI serve as an anchoring point for the next soliton. This conclusion is also supported by the fact that we never observed two or more L-pulses following each other. The formation of a patterned elastic soliton crystal from several H-pulse solitons and pairs of H-L pulses was observed. Thus, the proposed simple model embraces all main features of elastic soliton crystals observed in our experiment.

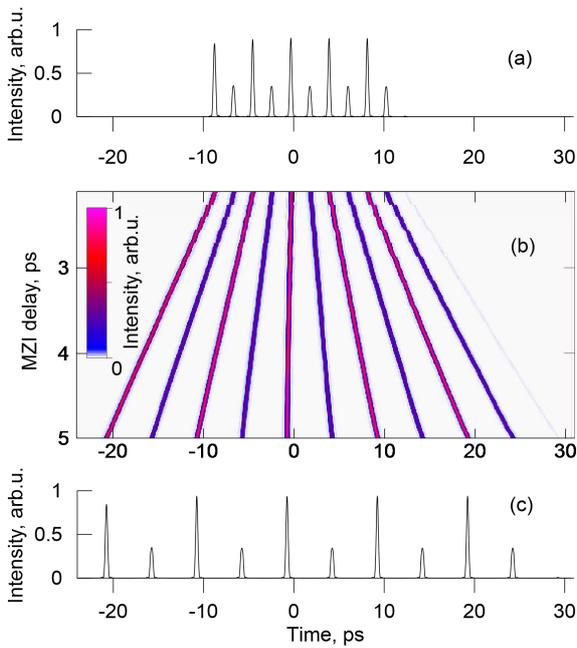

Fig. 8. Compression of soliton crystal comprising 5 pulse pairs: (a) intensity profile after compression, evolution of intensity profile during compression (b), intensity profile before compression (c).

## 6. DISCUSSION

We believe that our experimental findings supported by numerical simulations will help to better understand the processes underlying the generation of bound states of dissipative solitons and to introduce the novel concept of elastic soliton crystal.

The concept of elastic dissipative soliton crystals based on the new mechanism of long-range pulse-to-pulse interaction is a significant extension of the soliton crystal concept highlighting unusual properties that may have a notable impact on the fundamental studies of nonlinear dynamics in dissipative systems as well as on practical applications.

Mode-locked lasers implementing the concept of elastic soliton crystal may be highly important for practical applications. The generation of pulse bunches with easily and precisely tunable temporal separation is crucial for creating time fiducials, required, e.g., for synchronization of large experimental setups. Minimum spacing between pulses in the bunch limited in principle only by the pulse duration inside the cavity (which can be as small as a few 100 fs) may reach values corresponding to effective frequency in THz range. The repetition rate of the bunches can be controlled as well by changing the total cavity length. Although our laser was based on non-PM fibers, we believe that a fully PM scheme can be constructed based on PM NALM or graphene and topological insulator saturable absorbers with low saturation power [33].

The generation of continuous pulse trains with tunable ultrahigh repetition rate is of great interest for high-speed communications, frequency comb synthesis and various radiophotonic devices. Controllable excitation of a required high-order harmonic of the fundamental cavity frequency and intrinsic stability of the dissipative soliton parameters make the elastic soliton crystal laser a good candidate for the above applications.

The generation of stable ultrashort pulse patterns, which can be stretched and compressed in time conserving their structure, provide unique features for all-optical data processing and storage.

Although this was not tested in experiment, based on the observation of a large variety of the generated patterns we believe that almost any sequence of solitons and pairs can be generated in a laser cavity. We anticipate the possibility of on-demand pattern generation by utilizing external control of the pulse-to-pulse bond via placing an electro-optical amplitude modulator into one arm of the MZI. All-optical control is also possible by utilizing, e.g. Kerr gate as a modulator, or direct injection of pulses from external source into the cavity. This offers a promising way for creating optical and electrical signal pattern generators with sub-THz and even THz effective repetition rate. The pulse pattern formed with large spacing between pulses, which allows easy observation by standard oscilloscopes, can be then compressed down to short duration, providing a tunable ultra-high-speed source of the data patterns. On the other hand, the patterned elastic soliton crystal can be used as an intermediate ultrafast storage in all-optical data acquisition systems, in which the recorded pulse pattern then can be stretched to be measured by a standard oscilloscope.

While possessing some properties common to conventional soliton crystals, such as structural stability, self-assembling from fragments and dissociation into fragments under large perturbations, elastic soliton crystals have some unique features.

First, the "long-range" interaction induced by MZI instead of the "short-range" interaction induced by pulse wings ensures that elastic soliton crystals behave as a solid well-ordered structure while the pulse separation can be stretched to values much larger than the pulse size. This feature distinguishes elastic SCs from conventional SCs and real-world solid structures which have a long correlated order but do not demonstrate such enormous elasticity.

Second, in the elastic soliton crystals the bond formation mechanism supports coexistence of at least two different basic structures. This enables formation of patterned elastic soliton crystals.

Third, a specific feature of the elastic soliton crystals is nonreciprocal nature of the pulse-to-pulse interaction. Each pulse in the sequence influences significantly only the next pulse, but not the previous one, given positive delay of the small replica in the MZI. Thus, any excitation or local deformation of the elastic soliton crystal from ideal shape can propagate mainly in one direction. This may lead to interesting and unusual collective dynamics which is to be investigated in future works.

In conclusion, we have proposed the concept of elastic optical soliton crystals, which relies on a long-range pulse-to-pulse interaction in a mode-locked laser with intracavity Mach-Zehnder interferometer. By utilizing optical sampling we directly observed generation of stable and robust soliton crystals with tunable pulse separation in the fiber laser with intracavity MZI. We demonstrated that soliton crystal can be formed by two types of primitive structures: single dissipative solitons, and(or) pairs of dissipative soliton and pulse with lower amplitude. We demonstrated continuous stretching and compression by a factor of more than 30 of soliton crystals with conserved structure. In support of the observed experimental results we performed numerical simulation of a model laser cavity that reproduces the main properties of elastic soliton crystals. We studied in experiment collective pulse dynamics, including soliton crystal self-assembling, cracking and transformation of crystals comprising pulse pairs to the crystals of similar pulses.

**Funding.** Russian Science Foundation (RSF) (17-72-10236).

See Supplement 1 for supporting content.

# Extremely elastic soliton crystals generated in a passively mode-locked tunable high-repetition-rate fiber laser: supplementary material

ALEXEY ANDRIANOV,[*] ARKADY KIM

*Institute of Applied Physics of the Russian Academy of Sciences, Nizhny Novgorod, Russia*
*\*Corresponding author: **alex.v.andrianov@gmail.com***

This document provides supplementary information to "Extremely elastic soliton crystals generated in a passively mode-locked tunable high-repetition-rate fiber laser"

**Numerical model of passively mode-locked laser incorporating Mach-Zehnder interferometer (MZI).** Our modeling was primarily intended to show the possibility of formation of elastic soliton crystals and demonstrate their basic properties rather than precisely describe the pulse evolution in our experimental laser cavity.

To simplify the model we simulated a ring cavity instead of a figure-of-eight cavity. We did not directly model the propagation of the pulse inside the nonlinear amplifying loop mirror (NALM), instead we replaced it by an artificial saturable absorber with model transfer function. However, dispersion, Kerr nonlinearity and gain of the fibers inside the NALM were deliberately included into the model of the propagation loop. We did not take into account the evolution of the polarization state either.

The model cavity included three main elements: an amplifier with dispersion and Kerr nonlinearity, a saturable absorber and an MZI. The effects of dispersion, Kerr nonlinearity and gain of all fibers in the cavity were modeled by one element, in which the pulse propagation was described by the following equation

$$\frac{\partial A}{\partial z} = \left(\frac{g}{2\Omega^2} - i\frac{\beta_2}{2}\right)\frac{\partial^2 A}{\partial t^2} + \frac{g-\alpha}{2}A + i\gamma |A|^2 A \quad \textbf{(S1)}$$

Here, $\beta_2$ is second order dispersion coefficient, $\Omega$ is gain bandwidth, $\alpha$ is fiber loss, $\gamma$ is nonlinearity, and $g$ is gain coefficient.

Assuming that the population inversion of active ions doesn't change significantly during one cavity round-trip time $T_c$, slow gain dynamics can be modeled by [1]

$$g = g_0\left(1 - \frac{P_{av}}{P_s}\right)^{-1} \quad \textbf{(S2)}$$

where $\tau$ is "slow" time, $\tau=nT_c$, n is round-trip number, $g_0$ is unsaturated gain, $T$ is relaxation time, $E_s$ is saturation energy, and $E$ is the energy of the pulses in the cavity.

Equation S1 can be numerically integrated by the split-step method. To further speed up the modeling we simulated the propagation over the whole length $L_c$ in only two fast steps: the first one takes into account dispersion and gain, and the second Kerr nonlinearity.

The propagation through the saturable absorber is modeled by multiplying the field by the transmission function $f(A)$

$$f(A) = f_u + (f_s - f_u)(1 - \cos(|A|^2/P_{sa})) \quad \textbf{(S3)}$$

Here, $f_u$, $f_s$ are unsaturated and saturated transmission coefficients, respectively, and $P_{sa}$ is the saturation power of the absorber. Oscillating dependence of the transmission on the signal power of real NALMs is usually washed out at higher signal powers, however, in our simulation the value of $|A|^2$ did not exceed ~$3P_{sa}$, so that the periodic function (Eq. S3) can be used.

The MZI is modeled by splitting the signal into two parts and combining it with appropriate delay and phase. The signal at the MZI output is

$$A_{MZI}(t) = \sqrt{a}A(t) + \sqrt{b}A(t+T_{MZI})\exp(i\omega_0 T_{MZI}) \quad \textbf{(S4)}$$

Here, a and b are the transmission coefficients of the main and delayed pulses in the MZI, respectively, $\omega_0$ is signal central frequency, and $T_{MZI}$ is the unbalanced delay between two arms of MZI.

In the simulation the signal repeatedly propagated though the cavity and the result of the n-th cavity propagation cycle served as initial conditions for n+1 cycle. In our numerical code we implemented instant drawing of the current simulation result in real time and the possibility of manually adjusting the model parameters (the most important is MZI delay) during the simulation.

We used the following parameters in the simulation: $L_c$=7.8 m, $\beta_2$=-3 ps$^2$/km, $\gamma$=1W$^{-1}$km$^{-1}$, $T_c$=40 ns (although the simulation grid temporal span was 61 ps with 15 fs resoluton), $\alpha$=0.0003cm$^{-1}$, $E_s$=10 nJ (we took smaller value than real Es for Er-doped amplifier to speed up modeled gain dynamics), $P_s$ = 60 W, $f_u$=0.3, $f_s$=1, $T_{MZI}$ varied in the range 0-6 ps, and the ratio of the delayed pulse replica injected back into the cavity to the main pulse $b/a$ was varied in the range 0-0.05.